\documentclass[prb,preprint]{revtex4-1} 

\usepackage[english]{babel}
\usepackage[utf8]{inputenc}
\usepackage{graphicx}
\usepackage{subcaption}
\usepackage{caption}
\usepackage{amsmath}  
\usepackage{amsfonts} 
\usepackage{amssymb}
\usepackage{xcolor,soul}
\usepackage{physics}

\begin{document}

\title{Orbit Classification of asteroids using implementation of radial Basis Function on Support Vector Machines}
\author{Yashvir Tibrewal }
\affiliation{NPS International School, Singapore}
\author{Nishchal Dwivedi}
\email{nishchal.dwivedi@nmims.edu}
\affiliation{Department of Basic Science and Humanities, SVKM’s NMIMS Mukesh Patel School of Technology Management \& Engineering, Mumbai, India }

%\date{}

\begin{abstract}
This research paper focuses on the implementation of radial Basis Function (RBF) Support Vector Machines (SVM) for classifying asteroid orbits. Asteroids are important astronomical objects, and their orbits play a crucial role in understanding the dynamics of the solar system. The International Astronomical Union maintains data archives that provide a playground to experiment with various machine-learning techniques. In this study, we explore the application of RBF SVM algorithm to classify asteroids. The results show that the RBF SVM algorithm provides a good efficiency and accuracy to the dataset. We also analyze the impact of various parameters on the performance of the RBF SVM algorithm and present the optimal parameter settings. Our study highlights the importance of using machine learning techniques for classifying asteroid orbits and the effectiveness of the RBF SVM algorithm in this regard.

\end{abstract}

\maketitle
\section{Introduction}
Asteroid classification is a subject of immense importance for astronomical associations and governing bodies as potentially hazardous asteroids (PHA) present the potential to threaten global safety\cite{board2019finding}. Many attempts to do such classifications are in place \cite{barucci1987classification,klimczak2022comparison,tholen1989asteroid}. It is imperative that researchers and data scientists continue to update and develop the orbit classification models so that their accuracy can be increased and make precise predictions even with limits and uncertainties in the data. In addition to this asteroids give us insights about the past and present of our solar system \cite{michel2015asteroids}. Unaltered by the same forces that have changed Earth and larger planets these asteroids provide insights into the conditions of what the early solar system might have looked like and what factors resulted in the particular planetary formation we witness today. Our solar system is an especially interesting case study as star systems generally tend not to have more than 1 or 2 orbiting planets but the solar system has eight.

Asteroid classification plays an important role in planning and executing sky surveys, which are crucial for discovering and tracking asteroids \cite{ostro2002asteroid}. With more accurate and efficient asteroid classification algorithms, astronomers can plan better observing strategies, optimize telescope time, and increase the likelihood of detecting previously unknown asteroids.
Apart from their scientific significance, asteroids also have commercial potential. Asteroid mining is an emerging industry that aims to extract valuable minerals and resources from asteroids\cite{andrews2015defining}. Accurate asteroid classification is essential for identifying optimal candidates for mining and determining the stability of their orbits.

Classification algorithms are machine learning programs which perform analysis upon the input data \cite{mitchell2007machine}. They compare similar data points and classify them as groups in a process similar to feature engineering or pattern discovery. Numerous disciplines, including computer science, biology, marketing, and finance, use such algorithms. In addition to their use in asteroid classification, such algorithms can also be used to assist targeted advertising, genetic studies, and pattern recognition in financial data to guide investment choices.

Support vector machines (SVM) \cite{cortesc1995support} work by drawing a hyperplane which serves as a boundary and this boundary is maximised by considering the distance between points closest to the hyperplane. The radial Basis function (RBF) \cite{musavi1992training} kernal applies a nonlinear transformation to the data and projects the data in higher dimensions to increase the distance between the data points and draw a hyperplane which more accurately separates the classes making it the ideal choice for this classification problem. 

\section{Methodology}

The dataset used was the International Astronomical Union's minor planet centre (MPC) \cite{MPC} public access dataset which has been collecting data on asteroid orbit since 1995. The MPC identifies, designates, and determines the orbit of all minor planets and comets, including near-Earth objects and potentially hazardous asteroids. The dataset used for this study was extracted from the MPC's public access dataset on January 28th, 2023, and included data on over 800,000 asteroid orbits.

Once the data was obtained, it consisted of classified data of Main Belt Asteroids (MBA) and non-MBA asteroids (Refer Table \ref{asteroids}). The dataset was resampled such that the contribution from each of the type of the asteroid is equal and the dataset is not skewed. Only the asteroid types of Atira, Aten, Apollo and Amor can be potentially hazardous asteriods (PHA)\cite{ridpath2012dictionary,tedesco2005statistical,cuk2014hungaria}.

\begin{table}
\centering
\caption{Various types of Asteroids. The description is from the MPC dataset. $a$ is the semi-major axis. au is astronmical units.}
\label{asteroids}
\begin{tabular}{|p{2cm}|p{4cm}|p{10cm}|}
\hline
~ & Asteroid Type & Description \\
\hline
1 & Atira & ($a<1.0$ au $Q<0.983$ au)NEAs whose orbits are contained entirely with the orbit of the Earth \\
\hline
2 &Aten &($a<1.0$ au $Q>0.983$ au)	Earth-crossing NEAs with semi-major axes smaller than Earth's \\
\hline
3 & Apollo & ($a>1.0$ au $q<1.017$ au) Earth-crossing NEAs with semi-major axes larger than Earth's \\
\hline
4 & Amor & ($a>1.0$ au $1.017$ au$<q<1.3$ au) Earth-approaching NEAs with orbits exterior to Earth's but interior to Mars \\
\hline
5 & Hungaria & $1.78<a<2$au, $16< i<34$ \\
\hline
6 & Main belt asteroid MBA & $2.064<a<3.278$au, e,0.3\\
\hline
7 & Phocaea & $2<a<2.25$ au, $18<i<23$ \\
\hline
8 & Hilda &  roughly at 4 AU with $3:2$ Orbital resonance with Jupiter \\
\hline
9 & Jupiter Trojan & Located at Sun Jupiter lagrange points L4 and l5 with long term stable orbits \\
\hline
10 & Object with perihelion distance $< 1.665$ au  & -- \\
\hline
11 & Distant Object  & -- \\
\hline
12 & Unclassified & -- \\
\hline
\end{tabular}
\end{table}

\subsection{Support Vector Machines}
\begin{figure}[h]
\centering
\includegraphics[width=0.8\columnwidth]{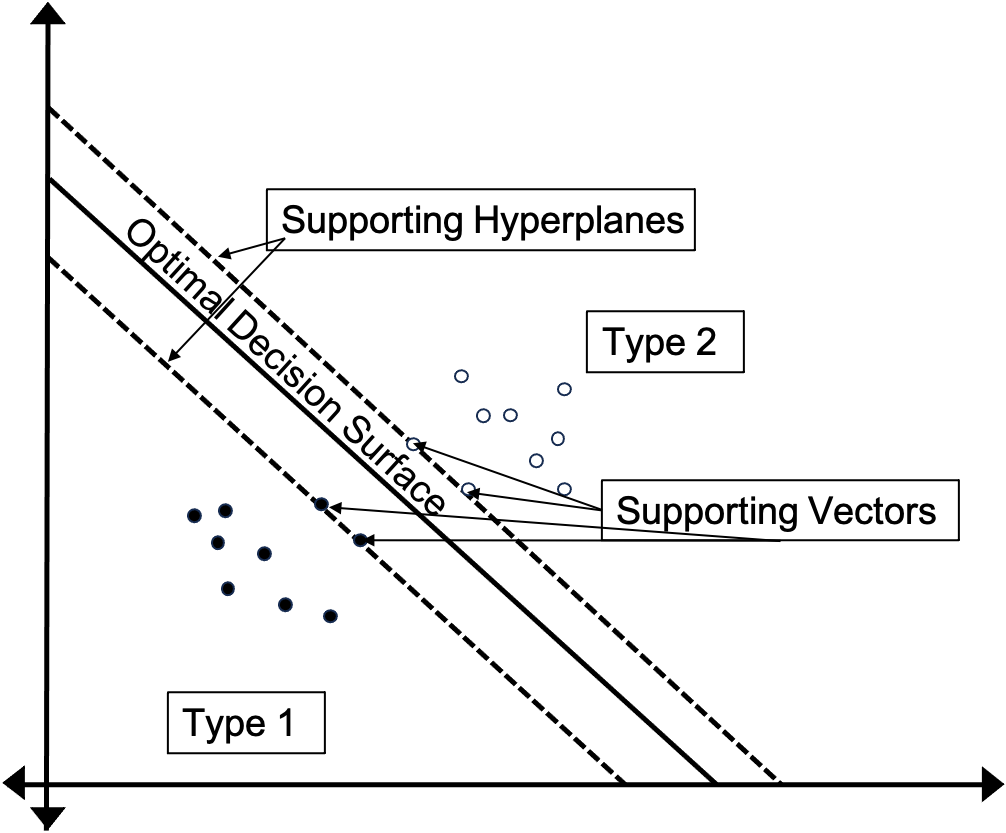}
\captionsetup{justification=centering} % center the caption
\caption{SVM Illustration} 
\label{SVM}
\end{figure}

Support Vector machines is a classification algorithm \cite{cortesc1995support} that creates decision boundaries by maximising the width of the margin. The margin width is determined by calculating perpendicular distance between the closest points from each class to the decision surface. These distances are also called support vectors. This can be visualised for linearly separable data as shown in the figure \ref{SVM}.

For a linear SVM, given a data set $(x_i,y_i)$, where $x_i's$ are a $p-$dimensional real vector. We try to find an optimal decision surface which divides the group of points into a clear classification. For a linear model this is an equation of the type $wx+b=0$. The closest data points to this hyperplane are then considered as support vectors and the distance between these points and the optimal decision surface is known as the margin. The algorithm attempts to find such a surface such that the margin is maximized. This leads to the maximum separation between the different groups in the data, and optimises the classification process\cite{cortesc1995support}.

The RBF kernel is a stationary kernel \cite{musavi1992training}. It is an exponential function which is infinitely differentiable. This kernel is used to convert a nonlinear dataset into a linear one. It adds an extra dimension to the problem using $z^2 = x_1 ^2 + x_2 ^2 + \ldots$, where $x_i's$ are the various variables of the problem. By increasing the dimension of the problem, it is easier to construct a hyperplane to separate various classes of classification. It relies on the fact that the position of the transformed data point is irrelevant only the relative distance from other points in the space is required to create the decision surface. 

It should be noted that several kernels that can be implemented such as the linear, polynomial and radial basis function (RBF). The linear kernel and polynomial project the data into higher dimensions by performing some linear polynomial transformation on the inputs to consider the correlations, where RBF uses an exponential function.

\section{Paramaters of RBF SVM}

There are two parameters which play a role in RBF SVM \cite{rbfparam}. The RBF kernel measures similarity between two points in the data as a Euclidean distance between them. The kernel function is defined as:
\begin{equation}
    K(x, y) = \exp\left( - \gamma ||x – y||^2 \right)
\end{equation}

where $x$ and $y$ are data points, and $||x – y||$ is the Euclidean distance between $x$ and $y$. $\gamma$ determines the width of the kernel. A larger $\gamma$ value leads to a more complex decision boundary which in turn is able to capture the finer nuances of the training data. Conversely, a smaller value of $\gamma$ results in a smoother decision boundary which can be more general for the new data, not allowing finer nuances to be captured \cite{rbfparam}.

For any learning algorithm, a generalisation error can be defined. This is a measure of accuracy of the algorithm to be able to predict correctly the outcome values for previously unseen data.

In RBF SVM, a hyperparamater called $C$ is used. This hyperparameter is used to control trade-off in the model between a good fit and a decision boundary. For larger values of $C$, a smaller margin of error is accepted, thus encouraging a more complex model. A lower $C$ will encourage a larger margin of errors. This would imply a a simpler decision function.

$\gamma$ and $C$ values impact the performance of the model significantly. These parameters are tuned for optimal values to find the best possible model for the given problem.

The table \ref{parameters} shows the optimisation of these parameters used.

The optimization of these parameters were done by creating a logspace for parameters $C$ and $\gamma$ which was used to develop a parameter grid which was randomly split using stratified shuffle split and then the highest accuracy score was returned along with the respective values of $C$ and $\gamma$. This parameter grid is rather large and due to computational limitations, the maximum grid size that could be implemented within reasonable computing time was $3\times3$. Therefore this process iterated 3 times with each iteration narrowing the parameter range and resulting in higher accuracy scores.

\begin{table}
\centering
\caption{Parameter optimization}
\label{parameters}
\begin{tabular}{|c|c|c|c|c|c|}
\hline
Iteration & C range  & Optimum C             & $\gamma$ Range & Optimum $\gamma$                                 & Accuracy \\
\hline
1         & (-2,9,3) & $10^{5}$ & (-9,2,3)    & $10^{-7}$                                          & 0.964    \\
\hline
2         & (4,6,3)  & $10^{6}$ & (-11,-7,3)  & $10^{-7}$                                          & 0.976    \\
\hline
3         & (6,7, 5) & $3.1\times 10^6$    & (-7,-8, 5)  & $3.16\times 10^{-7}$ & 0.979   \\
\hline
\end{tabular}
\end{table}

%%%%%%%%%%%%%%
\section{Implementation}
The features selected for the classification are given in Table \ref{features}.

\begin{table}
\centering
\caption{Various features which were selected for the implementation of the RBF SVM model.}
\label{features}
\begin{tabular}{|p{4cm}|p{12cm}|}
\hline
Features & Description \\
\hline
H & An asteroid’s absolute magnitude is the visual magnitude an observer would record if the asteroid were placed 1 Astronomical Unit (au) away, and 1 au from the Sun and at a zero phase angle  \\
\hline
M (degrees) & Mean anomaly is the fraction of an elliptical orbit's period that has elapsed since the orbiting body passed periapsis \\
\hline
Peri  Argument of perihelion (degrees) & Angle in the orbit plane between the ascending node and the perihelion point. \\
\hline
Node Longitude of the ascending node (degrees) &  It is the angle from a specified reference direction, called the origin of longitude, to the direction of the ascending node, as measured in a specified reference plane \\
\hline
i (degrees) & Angle between the orbit plane and the ecliptic plane \\
\hline
$e$  & the orbital eccentricity of an astronomical object is a dimensionless parameter that determines the amount by which its orbit around another body deviates from a perfect circle\\
\hline
$a$ (au) & One half of the major axis of the elliptical orbit; also the mean distance from the Sun \\
\hline
Orbital period (years) &  The time it takes an orbiting body to make one complete revolution around the Sun \\
\hline
Perihelion distance (au) & The perihelion distance is an astronomical term that refers to the closest distance between a celestial object and the Sun during its orbit.  \\
\hline
Aphelion distance (au)  & The Aphelion distance is an astronomical term that refers to the largest distance between a celestial object and the Sun during its orbit. \\
\hline
Semilatus rectum  (au)  & The chord through a focus parallel to the conic section directrix of a conic section is called the latus rectum, and half this length is called the semilatus rectum \\
\hline
n Mean daily motion & n (degrees/day) \\
\hline
U (Uncertainty) & integer with values 0–9 \\
\hline
\end{tabular}
\end{table}

\begin{figure}[h!]
\centering
\includegraphics[width=0.8\columnwidth]{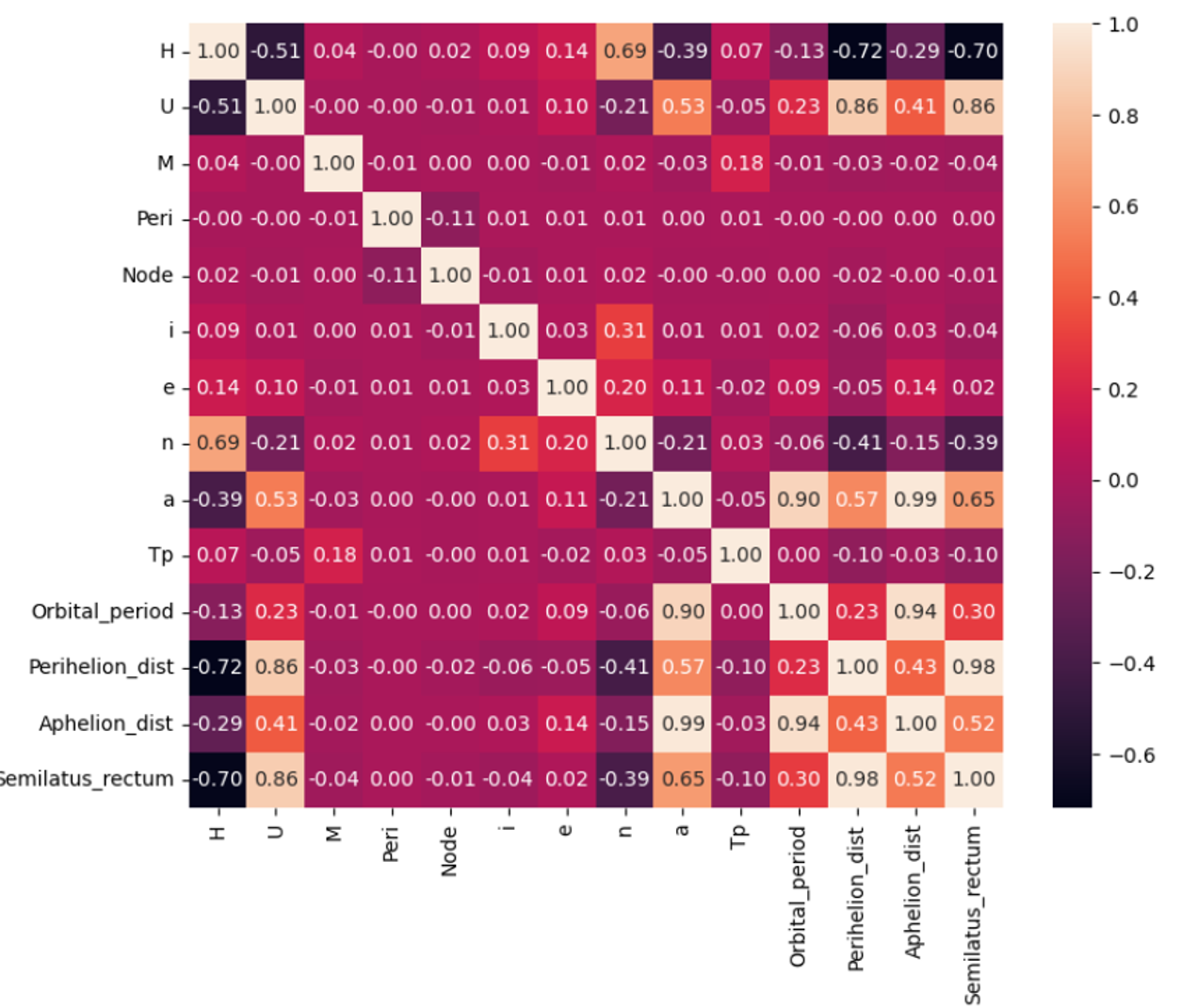}
\captionsetup{justification=centering} % center the caption
\caption{A Heat map showing corellations between the various features of the problem}
\label{heatmap}
\end{figure}

RBF SVM is implemented with the given data. The tuned parameters are used and a 80\% data is used for training and 20\% was used for testing. The RBF kernel was used to form hyperplanes, which seperated the kinds of asteroids successfully. The given figure \ref{heatmap} shows the correlation strength between any 2 variables. Most of the variables are independent of each other however there exist some inter-dependencies which cannot be avoided due to the nature of the problem. The orbital period is closely related to the perihelion distance and the aphelion distance as most orbital asteroids tend to have velocities within the range of 10 to 50 Km/s and therefore the distance of the orbit is directly correlated with the time taken to complete the orbit. The semi-latus rectum, aphelion distance and perihelion distance are all closely related to each other as most asteroids have eccentricities within the range of 0.05 to 0.35 with an average value of 0.17 thus resulting in a predictable generic pattern for these variables.

\begin{figure}[h!]
\centering
\includegraphics[width=0.8\columnwidth]{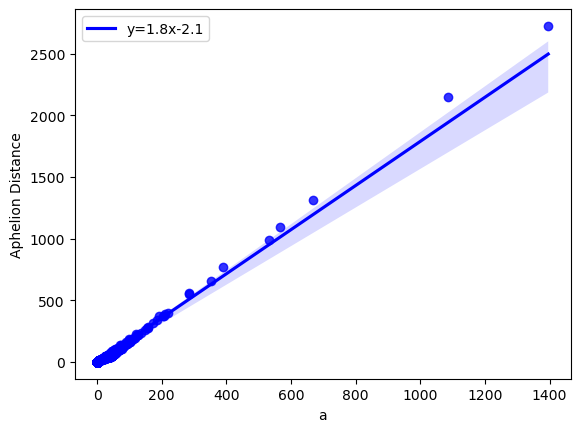}
\captionsetup{justification=centering} % center the caption
\caption{A Heat map showing corellations between the various features of the problem}
\label{apdist}
\end{figure}

The graph in figure \ref{apdist} depicts the closely related linear relationship of the aphelion distance against the semi-major axis which results from a relatively narrow range of eccentricities that most asteroids obey resulting in a mostly consistent orbit pattern for the majority of asteroids

\begin{figure}[h!]
\centering
\includegraphics[width=0.8\columnwidth]{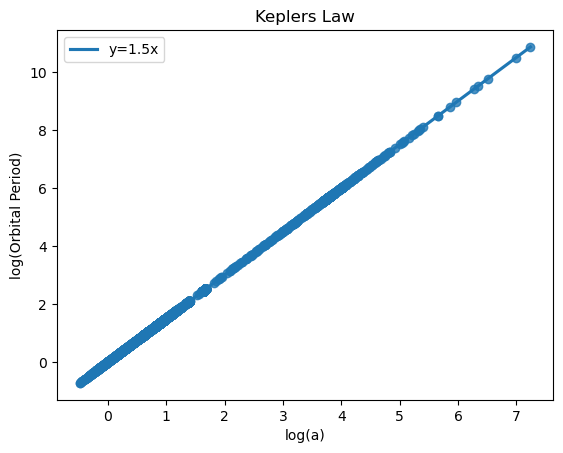}
\captionsetup{justification=centering} % center the caption
\caption{The log-log plot of Orbital time period and semi major axis for various asteroids. As expected, they follow the Kepler's third law, leading a slope of 1.5.}
\label{kepler}
\end{figure}

Figure \ref{kepler} verifies the Kepler's laws for the asteroids. The graph displays a clear and direct relationship between the orbital period and the semi-major axis of a planetary system, in line with Kepler's third law. The log-log plot of the orbital period and semi-major axis gives the slope of $\frac{3}{2}$ as shown in the figure \ref{kepler}.

\begin{figure}[h!]
\centering
\includegraphics[width=0.8\columnwidth]{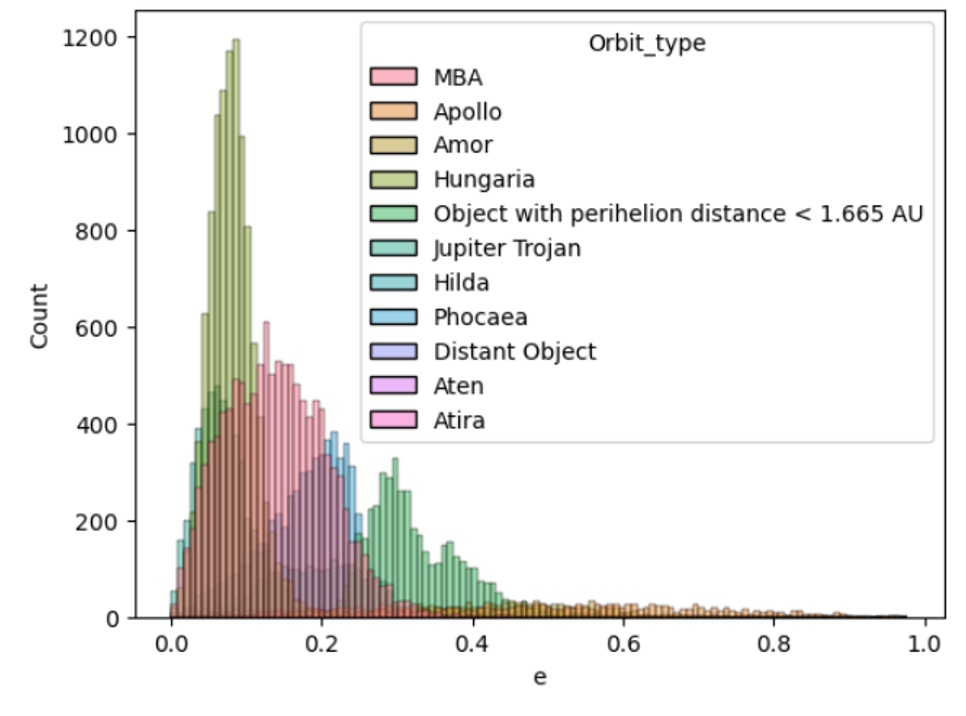}
\captionsetup{justification=centering} % center the caption
\caption{Histogram of eccentricities of various types of asteroids.}
\label{eec}
\end{figure}

The histogram in Figure \ref{eec} illustrates the various eccentricity values associated with different orbit types. The Hungaria group of asteroids are the innermost dense asteroids in the belt with semi-major axis lengths of 1.78 to 2 AU and therefore have a very narrow range of eccentricity closer to the orbit of a perfect circle. We also notice that Apollo asteroids have a very large range of eccentricities as they cross the orbit of larger planets such as Earth and have extremely large semi-major axes. 

\begin{figure}[h!]
\centering
\includegraphics[width=0.8\columnwidth]{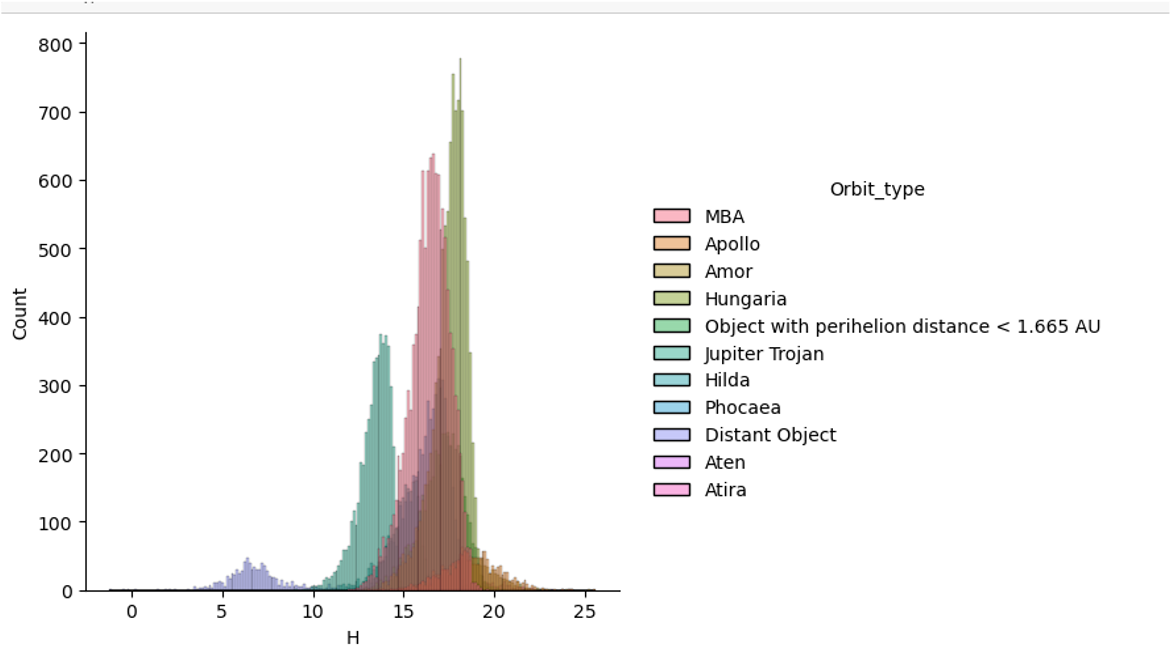}
\captionsetup{justification=centering} % center the caption
\caption{Histogram of the quantity H. There are clear distinctions in the peaks of H, which makes it a good variable for distinguishing the orbit types.}
\label{abs}
\end{figure}

\begin{figure}[h!]
\centering
\includegraphics[width=0.8\columnwidth]{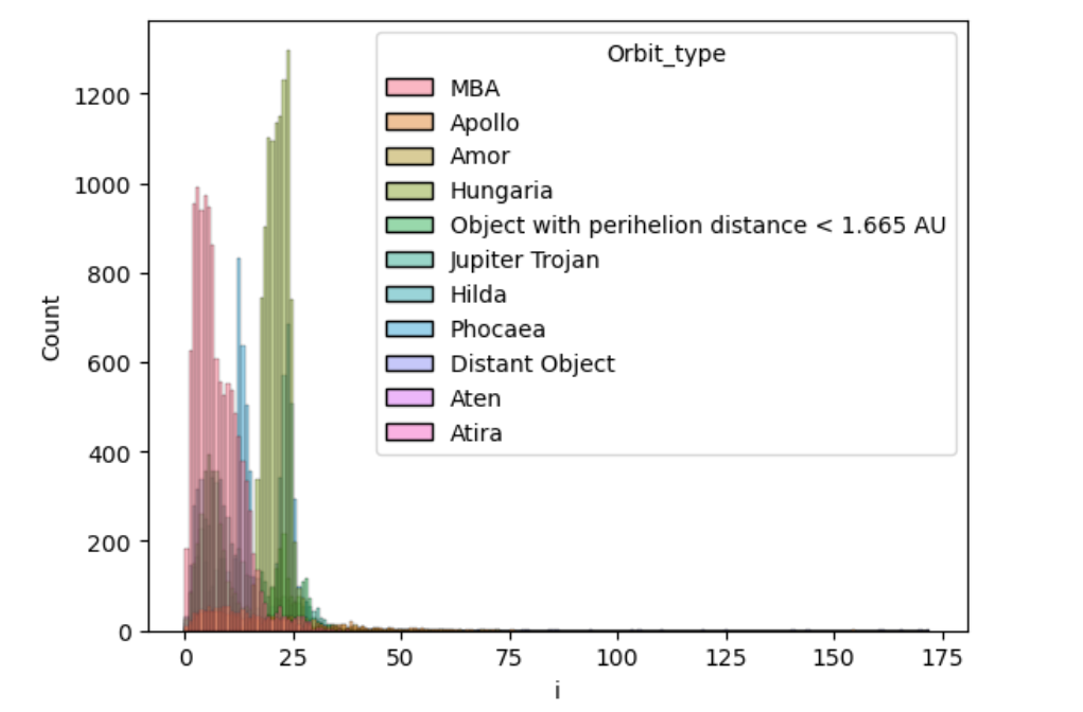}
\captionsetup{justification=centering} % center the caption
\caption{Histogram of the quantity i. There are clear distinctions in the peaks of i, which makes it a good variable for distinguishing the orbit types.}
\label{i}
\end{figure}

The inclination and Absolute magnitude also display some evident clusters for each specific orbit type, which can and will be exploited by the algorithm. These clusters result from some of the niche definitions that exist for some of these orbit types. For example, the Phocea family are a group of stony S-type composition asteroid located in the inner region of the asteroid belt. Their cluster can be seen peaking in the graph due to their exceptionally narrow range on inclinations from 18 to 23 degrees to the orbital plane. Similarly, the Jupiter Trojan asteroids have a distinct cluster on a scale ranging from 10 to 15. This results from the fact  absolute most Jupiter trojan asteroids have similar sizes and albedos. Jupiter trojan asteroids are a group of asteroids that share the same orbit as Jupiter, located at the L4 and L5 Lagrangian points. They are believed to have formed in the early Solar System, and their composition and physical properties are thought to be similar to those of the Kuiper Belt objects and the comets. These can be seen in Figures \ref{abs} and \ref{i}

\section{Results}

The RBF SVM code had a relatively high accuracy rate of 97.9\%. When compared this to other common classification algorithms \cite{klimczak2022comparison} and RBF SVM performs the best. Its success can be attributed to the capability of projecting the data to a higher dimension. 

\begin{figure}[h!]
\centering
\includegraphics[width=0.8\columnwidth]{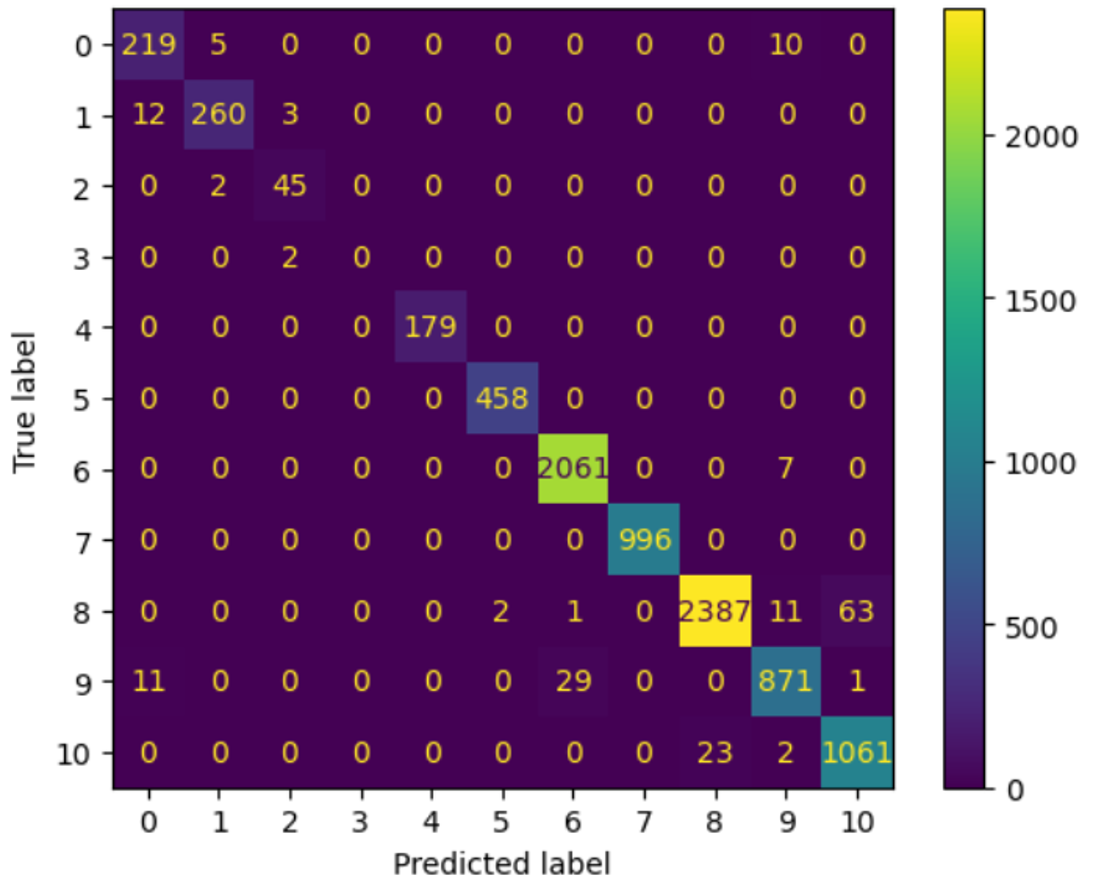}
\captionsetup{justification=centering} % center the caption
\caption{The confusion matrix gives the mis-classification of the orbits which belong to a particular type but are classified into another one. The numbers correspond to the orbit type from Table \ref{asteroids}. The dataset has around 9000 sample data.}
\label{confusion}
\end{figure}

The confusion matrix Figure \ref{confusion} provides a detailed overview of the performance of the RBF SVM algorithm in predicting the orbit types of asteroids. The numbers on the axes correspond to the type of orbit as given in order in Table \ref{asteroids} . The numbers along the horizontal axis each correspond with a predicted orbit type and along the vertical axis, the true orbit type, while the off-diagonal values represent the number of incorrectly predicted orbit types. As most of the values fall in the diagonal extending from the top left corner to the bottom right as the system makes predictions with high accuracy. The algorithm performs particularly well in predicting the orbit types that have a higher number of samples, such as types 1, 3, and 6, achieving accuracy rates of 95.7\%, 99.7\%, and 100\%, respectively. 

However, there are some pockets of inaccuracy where the algorithm confuses two orbit types, such as types 10 and 8, resulting in 63 errors. These errors may be due to the similarity between the features of these two orbit types, and further analysis may be needed to improve the classification performance. 

%In conclusion, RBF SVM has demonstrated significant promise over other classification algorithms in exploiting the higher dimensional patterns in the data and accurately drawing boundaries between orbit types. However, further research may be needed to address the limitations of the algorithm and improve its performance in accurately predicting the orbit types of asteroids.

\section{CONCLUSION}

In this research paper, we explored the classification of asteroid orbits using various clustering algorithms, with a specific focus on the radial Basis Function Support Vector Machines (RBF SVM) algorithm. This study highlights the importance of using machine learning techniques for classifying asteroid orbits and the effectiveness of the RBF SVM algorithm in this regard. 

The confusion matrix (Table \ref{confusion}) of the RBF SVM algorithm showed that it achieved high accuracy in predicting the orbit types of asteroids. The algorithm achieved an accuracy rate of 97.9\%

In conclusion, our study demonstrated the effectiveness of the RBF SVM algorithm in classifying asteroid orbits. This may have significant implications in understanding asteroids. By accurately predicting the orbit types of asteroids, we can improve our understanding of the dynamical and chaotic nature of star systems, and conditions of the early solar system and contribute to future space missions. However, further continued research is needed to properly address the limitations of our study and optimise the RBF support vector machines algorithm further. 

\section{Acknowledgements}

This research has made use of data provided by the International Astronomical Union's Minor Planet Center.

ND would like to thank Dr. Kumar Venkataramani (Caltech) for discussions.

\bibliographystyle{unsrt}
\bibliography{bib}
\end{document}